\documentclass[sn-mathphys]{sn-jnl}

\jyear{2021}%

\theoremstyle{thmstyleone}%
\theoremstyle{thmstyletwo}%

\theoremstyle{thmstylethree}%

\begin{document}

\title[Airport Digital Twins for Resilient Disaster Management Response]{Airport Digital Twins for Resilient Disaster Management Response}


\author*[1]{\fnm{Eva} \sur{Agapaki}}\email{agapakie@ufl.edu}

\affil*[1]{\orgdiv{M.E., Sr. Rinker School of Construction Management}, \orgname{University of Florida}, \orgaddress{\street{573 Newell Dr}, \city{Gainesville}, \postcode{32603}, \state{Florida}, \country{USA}}}


\abstract{Airports are constantly facing a variety of hazards and threats from natural disasters to cybersecurity attacks and airport stakeholders are confronted with making operational decisions under irregular conditions. We introduce the concept of the \emph{foundational twin}, which can serve as a resilient data platform, incorporating multiple data sources and enabling the interaction between an umbrella of twins. We then focus on providing data sources and metrics for each foundational twin, with an emphasis on the environmental airport twin for major US airports.}




\maketitle

\section{Introduction}\label{sec1}

\begin{figure}[h]%
\centering
\includegraphics[width=0.9\textwidth]{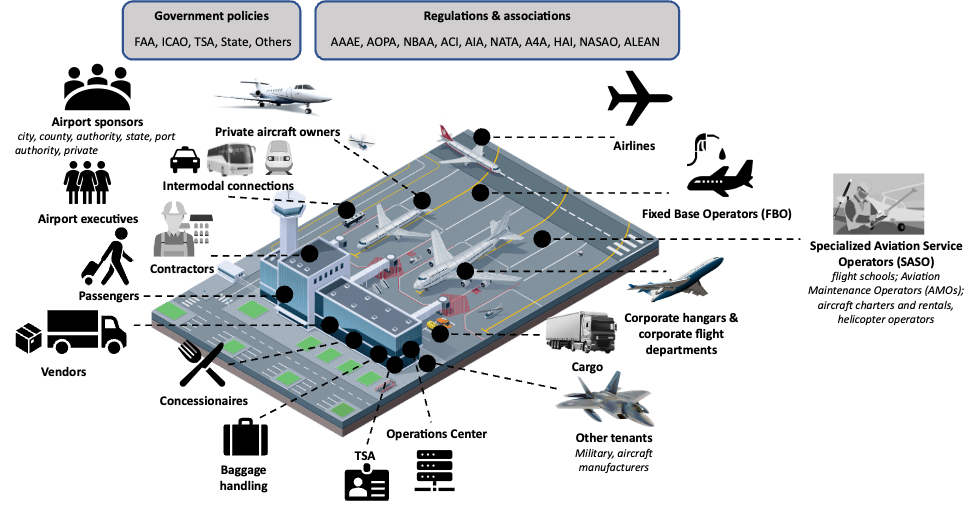}
\caption{Airport Systems and Stakeholders}\label{fig1}
\end{figure}

The complexity of airport operations regardless of their size extends beyond the airside side of operations. Natural disasters, climate change threats, high annual passenger demand, large volumes of cargo and baggage, concessionaires and vendors as well as other airport tenants may extend an airport’s operations beyond capacity or disrupt operations. Figure \ref{fig1} shows the airport systems and stakeholders in an airport. The American Society of Civil Engineers (ASCE) has rated airport infrastructure with grade “D” and this finding is based on the anticipated higher passenger demand compared to infrastructure capacity (Bureau of Transportation Statistics, 2019). Moreover, irregular operations and disruptions due to internal or external threats can have serious consequences in cities \citep{metzner2019comparison}. For example, in December 2017, one of the busiest airports in the world, Atlanta’s Hartsfield-Jackson airport, suffered an 11hr-long power outage, which disrupted the airport’s operations and also incurred economic losses \citep{sun2020resilience}. These incidents necessitate the need for a resilient airport.

Resilience incorporates the ability to (a) anticipate, prepare for, and adapt to changing conditions, (b) to absorb, (c) to withstand, respond to, and (d) to recover rapidly from disruptions. The implementation of resilient solutions in airports can be performed by preventing or mitigating disruptive events to air traffic operations \citep{clark2018resilience,yanjun2019measuring,pishdar2019influence}. Those events can be either severe weather hazards (e.g., dense fog, flooding, snow, drought, tornado, wildfire, hurricane), threats (e.g., equipment outages, political changes, economic downturn, pandemics, cyber-attacks, physical attacks) and vulnerabilities (e.g., equipment outages, lack of staff). 

Resilience can be quantified by analyzing risks to an airport. We adopt the definition of airport risks by the National Infrastructure Protection Plan (NIPP) \citep{NIPP}, where risk is defined by the likelihood and the associated consequences of an unexpected event. Those risks are the hazards most likely to occur, potential threats, and vulnerabilities. Hazards and threats refer to incidents that can damage, destroy, or disrupt a site or asset. The difference between hazards and threats is that the former can happen unexpectedly, typically outside of an airport’s control whereas the latter happen purposefully and are usually manmade. Some examples of hazards are natural hazards (e.g., hurricanes, earthquakes, wildfire), technological (e.g., infrastructure failure, poor workmanship, or design), or human-caused threats (e.g., accidents, cyberattacks, political upheaval). The consequences associated with the vulnerabilities of an airport, as a result of a hazard or threat being realized, is one way to measure the impacts associated with risks. Therefore, risk is defined in \ref{eq1} by:

\begin{equation}
\textrm{Risk} = \textrm{Consequence} \quad x \quad \text{probability} \quad x \quad \textrm{vulnerability}.\label{eq1}
\end{equation}

\begin{figure}[h]%
\centering
\includegraphics[width=0.9\textwidth]{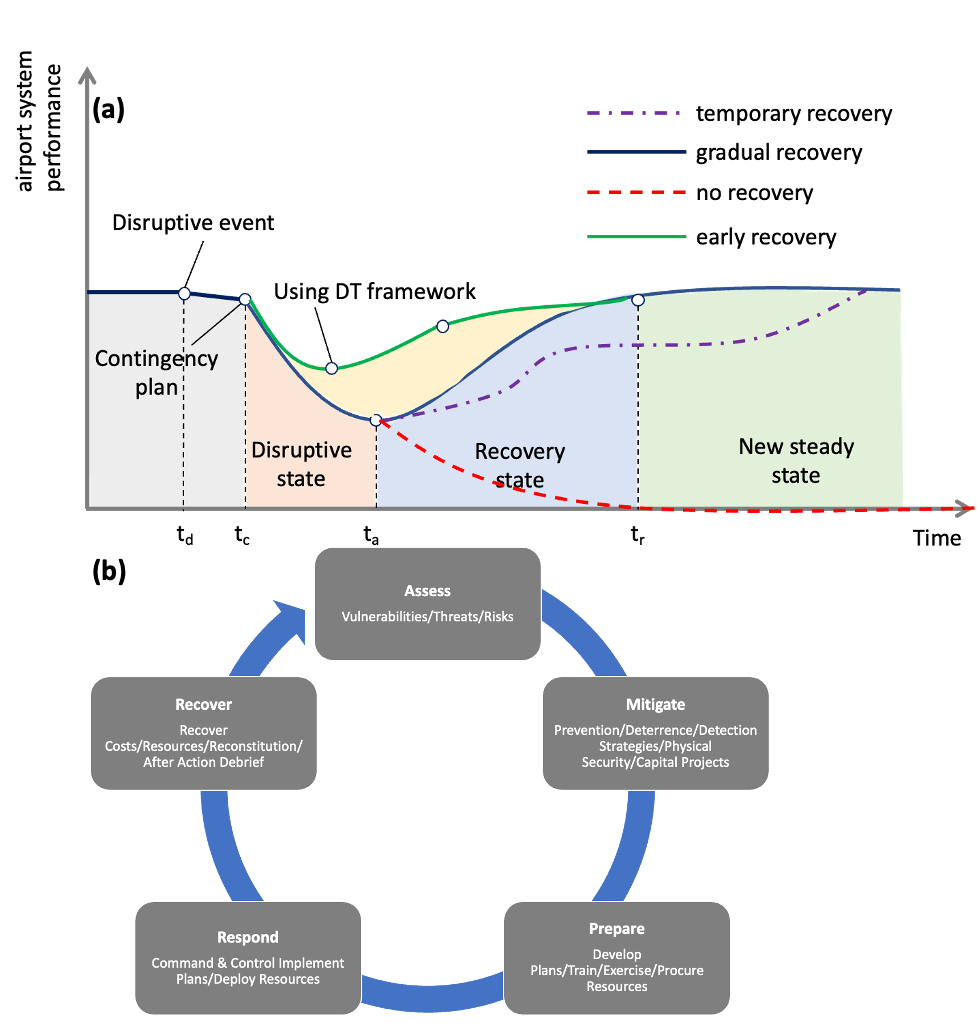}
\caption{(a) Resilience framework overview with and without the use of Digital Twins and (b) risk assessment adoption in airports (modified from Crosby et al., 2020).}\label{fig2}
\end{figure}

Resilience analysis includes both the time before (planning capability), during (absorbing capability) and after a disruption event occurs (recovery and adaptation capability), including the actions taken to minimize the system damage or degradation, and the steps taken to build the system back stronger than before. Figure \ref{fig2}(a) shows this timeline and the planning \citep{yang2015key,huizer2015usefulness,humphries2015evaluation,skorupski2016fuzzy,chen2016safety,zhao2017application,singh2019investigating,ergun2019assessment}, absorbing \citep{yang2015key,skorupski2016fuzzy,chen2016safety,zhao2017application,tahmasebi2018integrated,willemsen2018extending,singh2019investigating,ergun2019assessment}, recovering \citep{yang2015key,zhao2017application,wallace2017disaster,zhou2018emergency,bao2018measurement}, and adapting \citep{yang2015key,chen2016safety,zhao2017application,wallace2017disaster,zhou2018emergency,ergun2019assessment} phases of a resilience event. As shown in Figure \ref{fig2}, the system initially is in a steady state. After the disruptive event occurs at $t_d$, the system’s performance starts decreasing and then a contingency plan is implementing at time $t_c$. Then, there are four “recovery” scenarios. In the first scenario (blue line), the performance of the system gradually recovers without any outside intervention until it reaches the original steady state. In the second scenario (purple, dashed line), the system first reaches a new steady state, but eventually returns to its originally state. For example, temporary routes and measures are taken to meet immediate needs of airport operations when the system is damaged due to a hurricane. However, it may take weeks or months for the system to fully recover. The worst scenario is when the system cannot recover (red dashed line). The last scenario is to reach the recovery state earlier by using a holistic Digital Twin (DT) framework (green line), which will be discussed in the last section of this paper. Figure \ref{fig2}(b) showcases the risk assessment adoption framework in airport operations \citep{crosby2020}.

This paper targets to identify the areas of highest risks for an airport, so that these can serve as indicators to inform policies and investment decisions.

\subsection{Background on airport resilience}

In recent years, there has been a lot of research on resilience in airport operations. Multiple studies have implemented Cost-Benefit Analysis (CBA) tools to estimate costs and benefits after implementing security measures in their security risk assessment policies. However, CBA analysis cannot validate the estimated airport security costs, therefore multiple simulation experiments are needed to investigate the interdependencies between different stakeholders and systems \citep{stewart2014cost}. To overcome these limitations, the ATHENA project investigated a framework to evaluate curbside traffic management control measures and traffic scenarios at the Dallas-Fort Worth International Airport (DFW) \citep{UGIRUMURERA2021130}, optimization of shuttle operations that can lead to 20\% energy reductions \citep{SIGLER2021102077} and traffic demand forecasting \citep{LUNACEK2021102061}. 

Researchers have also focused on risk assessment models of airport security. Lykou et al. (2019) \citep{lykou2018smart} developed a model for smart airport network security with the objective to mitigate malicious cyberattacks and threats. Zhou and Chen (2020) \citep{zhou2020measuring} proposed a method to evaluate an airport’s resilient performance under extreme weather events. Their results demonstrated that airport resilience greatly varies based on the level of modal substitution, airport capacity and weather conditions. Previous research has greatly focused on airport security protection \citep{yanjun2019measuring,zhou2019resilience,thompson2019operational}. Agent-based modeling has been used to represent sociotechnical elements of an airport’s security system and identify states and behaviors of its agents such as weather, pilots, aircrafts, control tower operators \citep{stroeve2017agent}. Recently, Huang et al. (2021) \citep{huang2021building} proposed a Bayesian Best Worst Method that identifies the optimal criteria weights with a modified Preference Ranking Organization method for Enrichment evaluations (modified PROMETHEE) to make pairwise comparisons between alternatives for each criterion. They evaluated their method in three airports in Taiwan. This system relies on the judgement of experts for the evaluation of multiple, even overlapping criteria based on pre-determined evaluation scales.

However, a comprehensive framework for resilient management response for airports has not yet been developed. This is a complex and difficult Multiple Criteria Decision-Making (MCDM) problem. The objective of MCDM is to identify an optimal solution when taking into account multiple overlapping or conflicting criteria. This study intends to develop a framework that leverages recent advances in digital twin technologies and identify the criteria and metrics, which will act as a guidance for a holistic disaster management response.

\subsection{Resilience Indexes}

\begin{figure}[H]%
\centering
\includegraphics[width=0.7\textwidth]{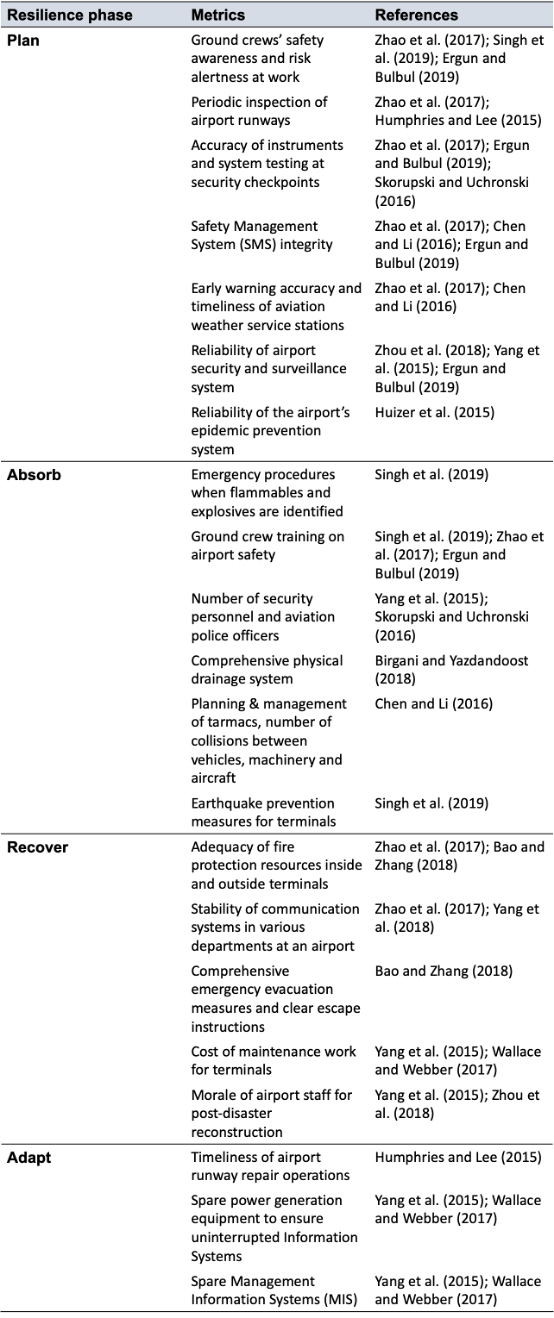}
\caption{Literature review on metrics for each airport resilience phase}\label{fig3}
\end{figure}

Previous literature has focused on identifying multiple metrics (resilience indexes) related to aviation and airport safety and operations \citep{bruneau2003framework,yang2015key,huizer2015usefulness,humphries2015evaluation,skorupski2016fuzzy,chen2016safety,zhao2017application,wallace2017disaster,wallace2017disaster,tahmasebi2018integrated,willemsen2018extending,zhou2018emergency,bao2018measurement,damgacioglu2018route,janic2019modeling,singh2019investigating,ergun2019assessment,metzner2019comparison}. Figure \ref{fig3} summarizes the most widely used resilience metrics for each resilience phase as described above. These metrics take into account the airport’s physical facilities, personnel, equipment and the disaster response phases.

Resilience indexes and web-based platforms have been widely developed for communities. Two widely adopted tools are the National Risk Index (FEMA, 2022) and the ASCE Hazard Tool (ASCE, 2022). The former calculates risk index scores per each US county based on 18 natural hazards by computing the expected annual loss due to natural disasters multiplied by the social vulnerability divided by the community resilience. The ASCE Hazard tool provides reports on natural disasters with widely known parameters. These tools need to be taken into account when considering the exposure of airports to natural disasters. 

The uncertainly of disruption occurrences in conjunction with the complexity of airport infrastructure and operations (Figure \ref{fig1}) necessitates the need for a unified digital platform to integrate information related to airport operations as well as interactions between airport sub-systems and their accurate representation.

\section{Airport Digital Twin Framework}\label{sec2}

Technological innovations have the potential to: (a) capture the detailed geometry of the physical infrastructure and generate the asset’s digital twin, (b) enrich the geometric digital twin with real-time sensor data, (c) update, maintain and communicate with the digital twin and (d) leverage the digital twin to monitor the asset’s performance and improve decision-making by planning interventions well before the time of need.
The goal of this research is to develop a foundational digital twin template that can be implemented across airports of all sizes. This leverages the use of digital twin technologies to explore alternative future scenarios for a more resilient airport. The foundational digital twin will be used as the main framework and specific systems of the digital twin will be investigated. In particular, the key objectives are the following:

\begin{itemize}
    \item {O1: Airport Digital Twin (ADT) definition in the context of operational airport systems.}
    \item {O2: Airport Digital Twin (ADT) generation and maintenance. We propose a framework for static and dynamic information curation based on existing sensor data and airport systems.}
\end{itemize}

\begin{figure}[h]%
\centering
\includegraphics[width=0.9\textwidth]{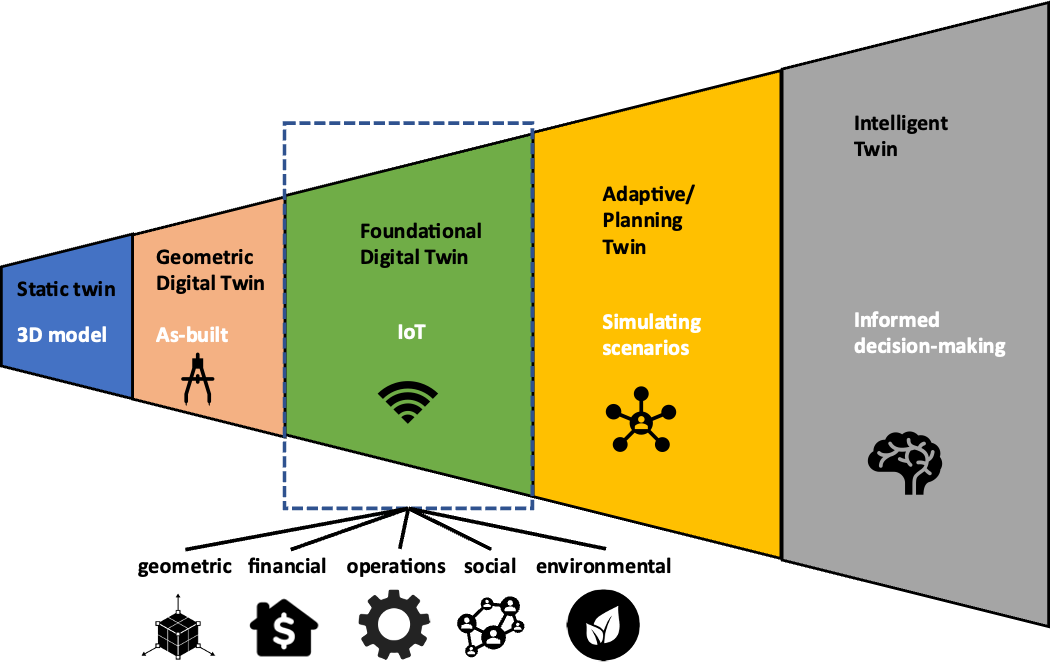}
\caption{Digital Twin framework}\label{fig4}
\end{figure}

We introduce the concept of the foundational airport digital twin (Figure \ref{fig4}), which incorporates an umbrella of twins that could interact with each other; these twins are, but not limited to, the geometric, financial, operations, social and environmental twin. If successful implementation of the foundational twin is achieved, then it can lead to improved efficiency and operations as well as better planning in the presence of irregular events that are of paramount importance for airport executives and stakeholders. The Foundational Digital Twin will serve as a resilient data backbone for airport infrastructure systems and will enable the implementation of more advanced twins such as the adaptive/planning and intelligent twin. As illustrated in Figure \ref{fig4}, the adaptive twin encapsulates simulated scenarios towards a proactive plan of operating an airport, where planned interventions will be more sophisticated than ever before. The data collection, modeling and intervention will become increasingly automated. That level of automation will lead to the Intelligent Twin, where we envision an informed decision-making system with minimal human intervention.

\begin{figure}[h]%
\centering
\includegraphics[width=\textwidth]{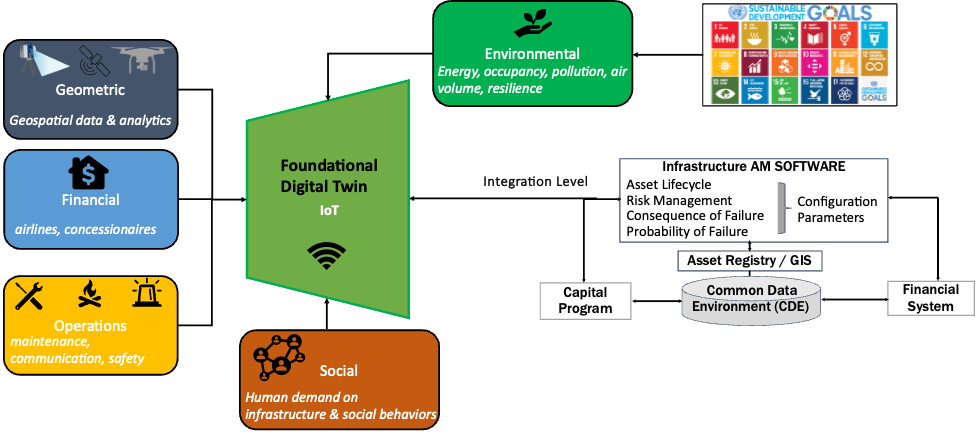}
\caption{Definition of foundational digital twin}\label{fig5}
\end{figure}

We expand on the definition of the Foundational Digital Twin in Figure \ref{fig5}. The geometric twin entails spatial data collection as well as their intelligent processing, Building Information Modeling (BIM) and validation with laser scanned data and GIS data integration. The process of laser scanning to BIM has already been applied to other complex infrastructure assets and is named as geometric digital twinning \citep{agapaki2020automated}. The Financial twin should have the capacity to: (a) simulate the allocation of funding from a variety of sources and visualize it in different physical assets at the airport, (b) visualize potential conflicts in funding utilization in real-time and (c) facilitate the fiscal management of an airport expansion and renewal projects. The Social twin should visualize the human demand on infrastructure and predict social behaviors based on historical data. In particular, it should: (a) integrate geospatial and airport-specific data (e.g., area and airport infrastructure reachability in correlation to the number of runways, taxiways), (b) integrate and process demographic data such as urban indexes and population around the airport to predict human demand and (c) integrate geographic and urban data related to the existing built environment surroundings that can affect passenger demand. Lastly, the environmental twin should account for natural hazards, energy consumption, occupancy rates, pollution and air volume for airports to be on track to achieve net-zero carbon infrastructure by 2050 (UN Environment Program, 2020).

As presented in Figure \ref{fig5}, the foundational digital twin will be integrated with existing infrastructure Asset Management (AM) software and the configuration parameters (asset lifecycle, risk management, consequence of failure, probability of failure) will be predicted based on the capabilities of the twin. The asset will be registered in a Common Data Environment (CDE) and the infrastructure needs will be assessed based on the existing capital program of the airport (in-year or multi-year capital program) as well as the financial system (e.g., existing asset reports, tangible capital asset and government accounting standards board). All the data is expected to be aggregated in a data warehouse (data lake), which will be hosted in the airports’ Operations Center facilities.

\subsection{Exploration of threats and hazards for Airport Digital Twins}

\begin{figure}[h]%
\centering
\includegraphics[width=\textwidth]{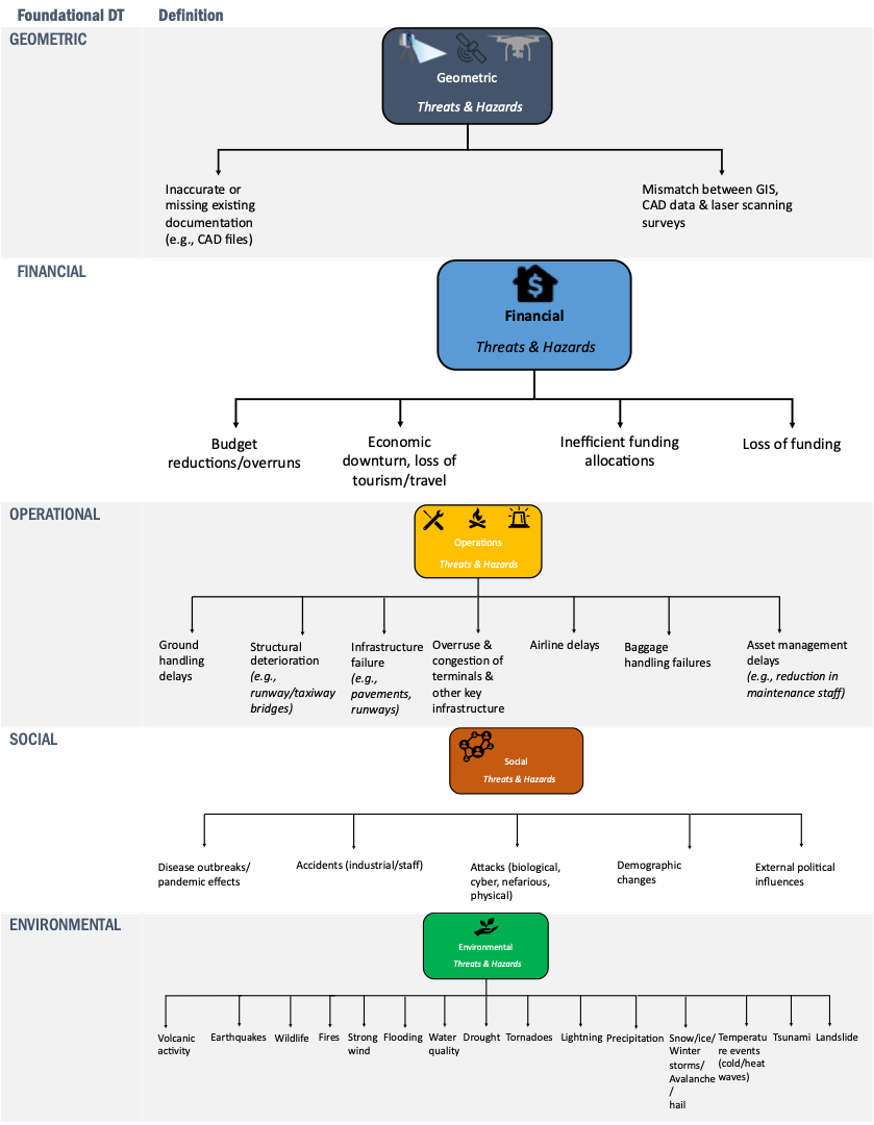}
\caption{Definition of Foundational DTs in relation to hazards and threats}\label{fig6}
\end{figure}

The threats and hazards are grouped into each Foundational Twin and are summarized in Figure \ref{fig6}. Each group is then analyzed below.

{\bf Geometric.} The majority of aging industrial facilities lacks accurate drawings and documentation \citep{agapaki2020automated}. Without capturing the existing geometry of an asset accurately, the incurred information loss throughout an asset's lifecycle would be immense.

{\bf Financial.} Budget reductions and overruns, economic downturns, inefficient funding allocations are some of the financial threats airport authorities may be encountered with \citep{graham2016airport,urazova2020infrastructure}. 

{\bf Operational.} Equipment maintenance is critical for airport operations (SMS Pilot Studies; FAA, 2019). The condition of passenger bridge boarding and ancillary equipment is another critical component that needs to be reliably assessed and maintained.

{\bf Social.} Airports are at risk of malicious events that can put their operations in jeopardy. Such incidents have been reported in the Los Angeles International airport in 2002 and 2013, when assailants killed airport personnel during attacks \cite{abdollah2013,Feldman2004}. In 2014 an intoxicated passenger attacked another passenger in a hate crime at the Dallas airport \citep{Kirkpatrick}, indicating the need for vigilance at all levels. Incidents of cyberattacks continue to increase and airport infrastructure is a target for those. Zoonotic diseases \footnote{A zoonotic disease is one in which an animal acts as intermediary for disease transmission between the vector and the infected human (e.g., Lyme disease occurs in white-footed mice, but it is transmitted to humans via ticks).}  will likely increase in their extent and frequency in the future \citep{mills2010}. We need to look into the diseases that have the highest likelihood of infecting regions where the airports operate. For example, in Texas, the most common vectors include ticks and mosquitoes carrying spotted fever rickettsiosis and West Nile virus, respectively \citep{CDC}. Other unknown zoonotic diseases could arise in the future in the same way that the novel coronavirus is thought to have emerged as a zoonotic disease and then rapidly spread around the world, primarily via airports and air travel \citep{CIDRAP2020}. In such scenarios, the airport’s operations could facilitate the spread of new zoonotic diseases from one person to another, including passengers, flight crews, airport workers, transportation personnel, and others. Other societal threats are demographic changes (e.g., sudden population growth) and unprecedented industrial or staff accidents. 

{\bf Environmental.} These threats include natural hazards and impacts of climate change that can significantly alter meteorological conditions, which may affect airport operations. Extreme natural events can lead to power system and flight disruptions. Studies have estimated that these events will be exacerbated to a modest degree by climate change \citep{Runkle2017,Hegewisch}. Although current research cannot definitively conclude whether climate change will increase or decrease the frequency of tornadoes in every situation, overall warmer temperatures will likely reduce the potential for wind shear conditions that lead to tornadoes \citep{Climate,Hausfather}. Severe wind could become more common because previous research has shown that climate change is responsible for a gradual increase in world average wind speeds \citep{ClimateCentral}. Fewer and/or less powerful tornadoes could help improve operations, because airport structures and visitors are less likely to be damaged or harmed by airborne debris caused by tornadoes. On the other hand, stronger overall winds could lead to more difficult weather conditions for pilots of arriving or departing aircraft. 

Climate change also has the potential to increase the annual average temperatures and to make the temperature swings more extreme. This can result in an increase in electricity demand as HVAC systems respond to temperature changes. This increased electricity demand can stress system components in both daily operations and in heat wave and cold snap events.

\begin{figure}[h]%
\centering
\includegraphics[width=\textwidth]{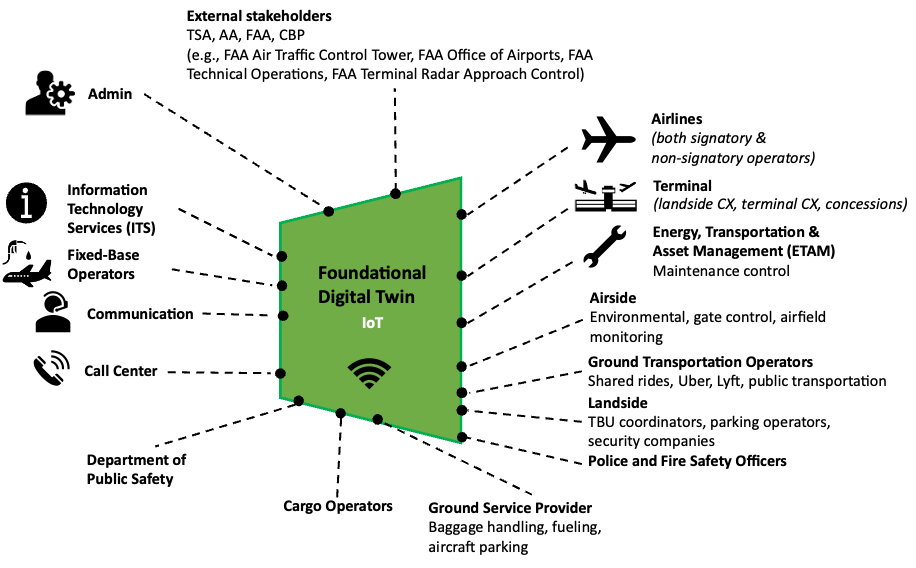}
\caption{Airport Stakeholders of the Operational Twin}\label{fig7}
\end{figure}

\begin{figure}[h]%
\centering
\includegraphics[width=\textwidth]{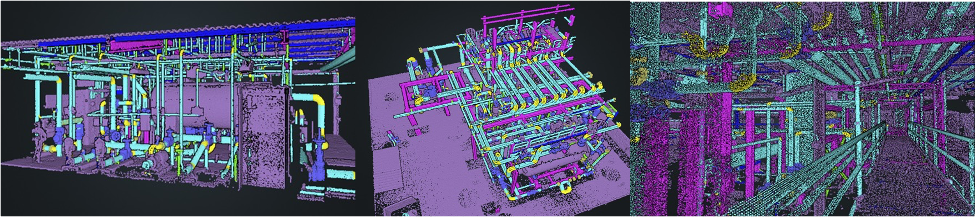}
\caption{Automated geometric digital twinning on industrial assets (Agapaki, 2020)}\label{fig8}
\end{figure}

There are three features of the twin that have the potential to assist decision making better than any other technology tool. Those are: (a) interoperability between software tools, which facilitates better communication between multiple stakeholders, (b) relationship mapping between both static (e.g., static infrastructure) and dynamic entities (e.g., humans, moving infrastructure) and (c) semantics that allow Artificial Intelligence (AI) tools and data analytics to forecast future scenarios. An example of airport stakeholders involved in the operational twin is presented in Figure \ref{fig7}. Figure \ref{fig8} shows semantics in the geometric digital twin applied to heavy industrial facilities. We will also investigate potential data sources for the generation and maintenance of DTs in the next section.

\subsection{Data sources for the generation and maintenance of airport DTs}

\begin{figure}[H]%
\centering
\includegraphics[width=\textwidth]{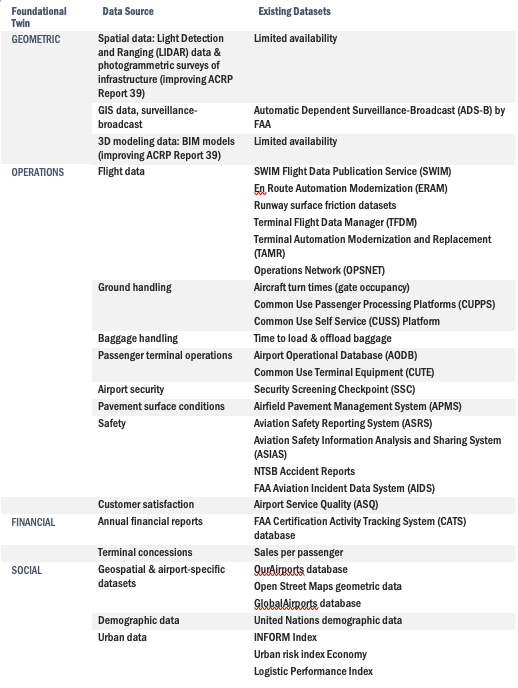}
\caption{Data sources and existing datasets for the geometric, operations,financial and social foundational twin}\label{fig9}
\end{figure}

\begin{figure}[H]%
\centering
\includegraphics[width=\textwidth]{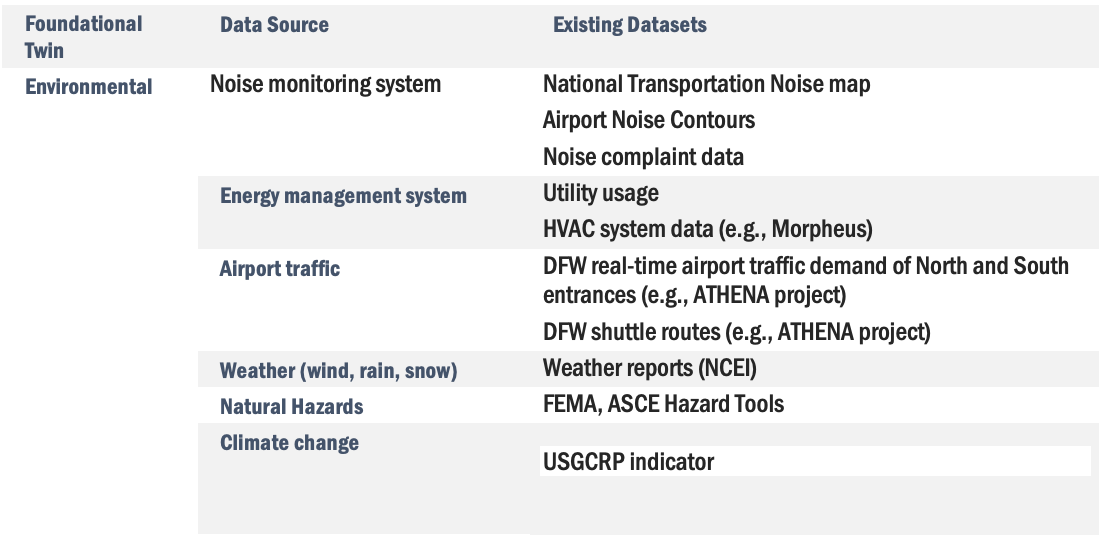}
\caption{Data sources and existing datasets for the environmental foundational twin}\label{fig10}
\end{figure}

A sample of data sources and existing datasets for each Foundational Twin is summarized in Figures \ref{fig9} and \ref{fig10}. Existing literature provides the insight that there is an abundance of systems related to air traffic control (i.e., ADS-B, SWIM, ERAM). FAA’s goal of the NEXT Generation Air Transportation System (NextGen) is a national upgrade of the air traffic control systems. The System Wide Integration Management (SWIM) is the first implementation of this vision, and it integrates a variety of data systems, including weather, communication radar information, traffic flow management systems, en route flight plan changes, arrival and departure procedures, microburst information, NOTAMs, storm cells, wind shear and terminal area winds aloft. However, the systems related to operating, maintaining and predicting failures and irregular operations in the landside and terminal operations are limited. Airport Computer Maintenance Management Systems (CMMS) have been proposed for use in airports (National Academies, 2015) to streamline maintenance operations by using work schedules, maintaining inventory and spare parts at optimal levels and tracking historical records. However, based on the results of the above-mentioned report, a total of 15 different CMMS software packages were used by airports that have implemented these systems, which makes the data governance and interoperability a challenge. In addition, although the exact location and tracking of air traffic is managed thoroughly through a variety of systems, the movement of passengers and occupants of the terminals is not monitored. To give an example, an airline becomes aware of a passenger being at the terminal only when they check in their luggage or pass TSA control.

The data needs will differ per airport size depending on their experience and on their operational management practices. At a minimum, we investigate data sources and metrics to be adopted for the environmental twin for a small set of US airports. 

\section{Investigation of environmental digital twin metrics}

We selected a set of airports, namely: Southwest Florida International airport (RSW), Hartsfield-Jackson International airport (ATL), Charlotte Douglas International airport (CLT), Ronald Reagan Washington International airport (DCA), William Hobby airport (HOU), Dalls/Fort Worth International airport (DFW), Dallas Love Field airport (DAL), Austin-Bergstrom International airport (AUS) and Orlando international airport (MCO). The reason for our selection was based on their number of enplanements, being in the top-10 busiest US airports for large and medium hub airports \citep{bazargan2003size}. 

\begin{figure}[H]%
\centering
\includegraphics[width=\textwidth]{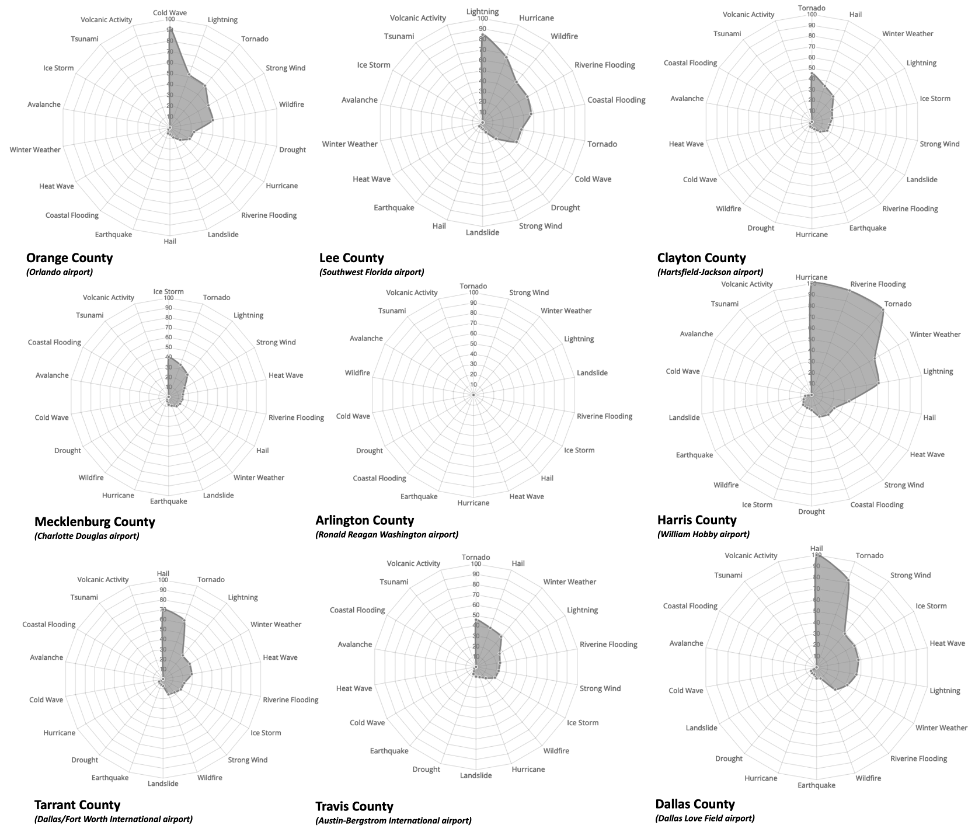}
\caption{Risk index scores for 18 natural hazards and nine US airports (modified from FEMA).}\label{fig11}
\end{figure}

We investigated the natural hazard risk index scores for selected counties, where the identified airports operate. Figure \ref{fig11} presents the relative distribution of hazard type risk index scores for each county, which can be used as one of the data sources for the environmental airport DT.

\begin{figure}[H]%
\centering
\includegraphics[width=\textwidth]{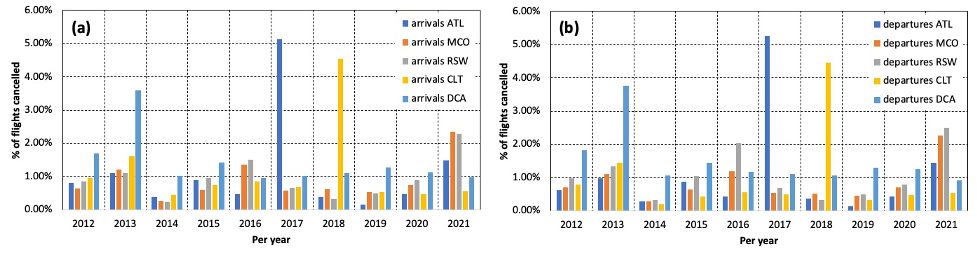}
\caption{Distribution of percentage of cancelled (a) arrival and (b) departure flights for ATL, MCO, RSW, CLT and DCA airports.}\label{fig12}
\end{figure}

\begin{figure}[H]%
\centering
\includegraphics[width=\textwidth]{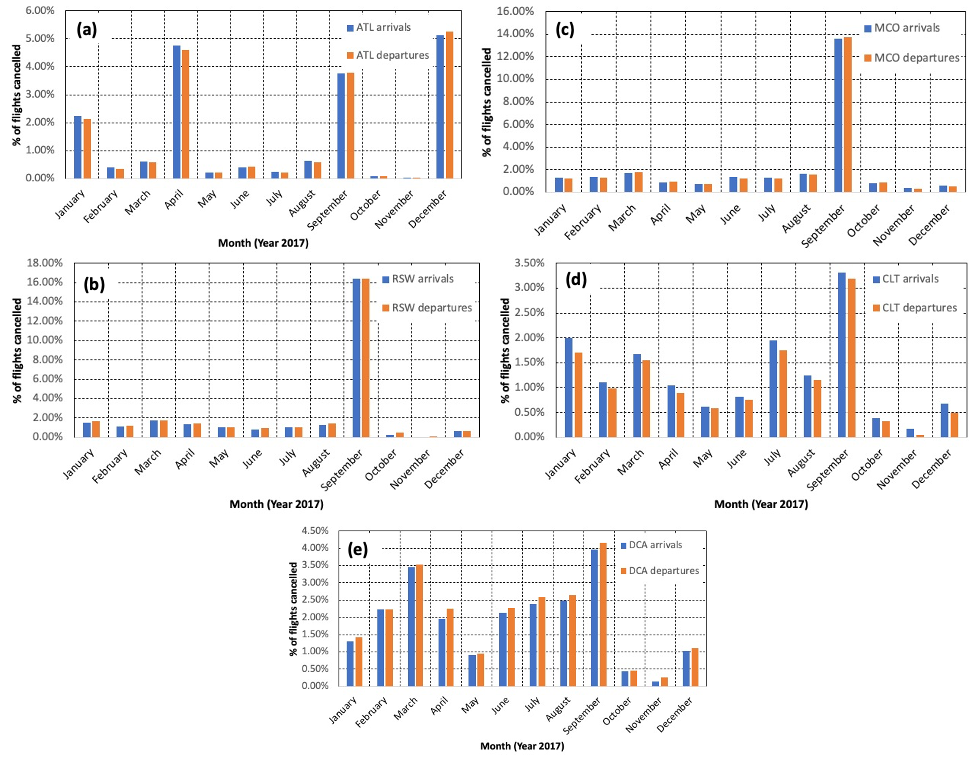}
\caption{Per month percentage of cancelled flights in 2017.}\label{fig13}
\end{figure}

Another indicator of irregular airport operations is the number of cancelled and/or delayed flights arriving or departing at the airport. We collected data from TranStats (Bureau of Transportation Statistics, 2022) for the on-time arrival and departure flights. Figure \ref{fig12} shows the arrivals and departures in five of the above-mentioned airports and their distribution per year for the last ten years. Then, we looked at extreme weather events that have occurred and could affect the operation of these airports. In particular, we investigated Hurricane Irma. Hurricane Irma occurred in September 2017 affecting MCO, RSW, ATL, CLT and DCA airports and primarily ATL airport with 3.76 and 3.78\% cancelled arrival and departure flights respectively. A power outage also affected the ATL airport with nearly 5\% cancelled flights in December 2017. Similarly, RSW, CLT, DCA and MCO were affected with the majority of their canceled flights being in September 2017 (Figure \ref{fig13}). Another important aspect that should be investigated is the recovery time of airport operations after these events occurred, which is part of future work.

Climate change metrics are another significant factor that needs to be taken into account when generating and maintaining a digital twin. The National Climate Assessment Representative Concentration Pathways (RCP) provide key indicators for quantifying climate change. Those are: cooling degree days, heating degree days, average daily mean temperature, annual number of days with minimum and maximum temperatures beyond threshold values, annual precipitation, dry days, and days with more than 2” precipitation. Cooling and heating degree days refer to the number of hours per year and the degrees above or below 65°F as detailed by the NOAA methodology (US Department of Commerce 2022). This methodology means that the cooling degree days and heating degree days can exceed 365 as they are multiplied with the number of hours per year and the magnitude of temperatures above or below 65°F. 

Temperature has the potential to impact human health. It can also affect power, transport, and water system resilience. For example, at the DFW airport, the number of cooling degree days is projected to increase from 2600 to 3400 and the number of heating degree days is projected to decrease from 2400 to 1900 (both according to the low emissions scenario). Cooling degree days will increase electricity consumption from air conditioning, increase on site water consumption, and can exacerbate the airport’s peak electricity demand which was analyzed in a subsequent section. Both maximum and minimum ambient temperature can affect electricity demand. The average high and low temperatures at 2 meters above the ground, the average precipitation and average wind speed are compared for ATL, CLT, RSW, DFW and HOU. The average precipitation is computed by accumulating rainfall over the course of a sliding 31-day period and the average wind speed is computed as the mean hourly wind speed at 10 meters above the ground.

\begin{figure}[H]%
\centering
\includegraphics[width=\textwidth]{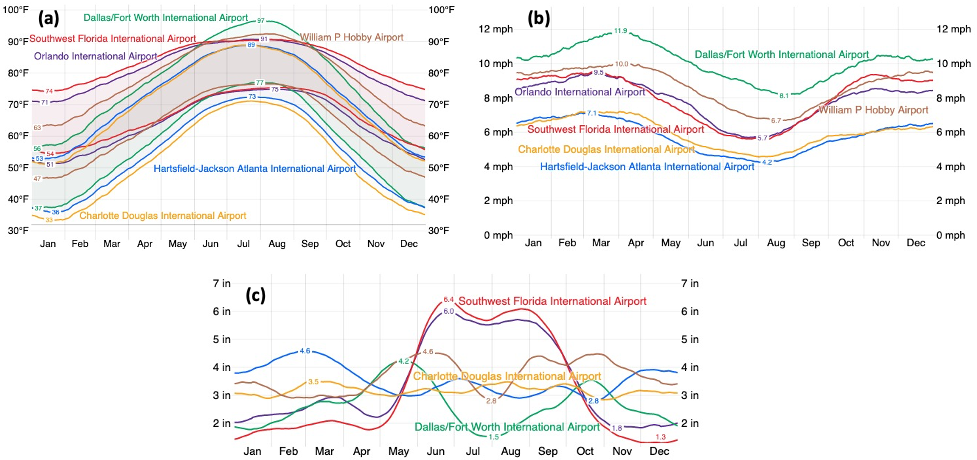}
\caption{(a) Average high and low temperature (b) average wind speed and (c) average monthly rainfall. }\label{fig14}
\end{figure}

Aside from the average temperature, wind and precipitation values, heat waves and cold spells can have an acute effect on airport operations. Using the SAFRAN methodology (EPA, 2021), those can be identified in future work.

The overall amount of precipitation, dry days, and days with significant rainfall ($>$ 2” per day) is expected to change very little from present time through 2050. We used DFW as an example, where there have been several historical events where extreme precipitation has led to power system and flight disruptions. These events have included snowfall (US Department of Commerce 2021a; US Department of Commerce 2021b; Narvekar 2011; NBCDFW 2021; Lindsey 2021; L’Heureux 2021), droughts (Runkle and Kunkle 2017; Hegewisch and Abatzoglou 2021; Centers for Disease Control and Prevention National Environmental Public Health Tracking 2020), and floods (USGCRP 2018; First Street Foundation 2021). 

\section{Discussion}\label{sec12}

When dealing with risk, an airport organization has the choice to either accept the risk, avoiding the risk by planning and preparing for interrupted activities or working to eliminate the risk through mitigation. This research reviewed the existing risks in the context of gathering available data sources, grouped them into categories (geometric, financial, social and environmental) and developed a unified framework for the assessment of risks using multiple criteria. We particularly emphasized environmental threats and provided metrics as well as open-source existing databases for the evaluation of those risks.

\section{Conclusion}\label{sec13}

There have been many studies investigating airport resilience frameworks, however a unified framework that identifies and then combines multiple data sources has not yet been investigated. The aim of this study was to use a foundational digital twins in order to identify key threats and hazards. We then suggested metrics and data sources for the environmental digital twin that can be used as guidance for the development of a unified and integrated data framework for the DT development. Future research directions include the foundational digital twin implementation on airport case studies.


\bibliography{sn-bibliography}


\end{document}